\documentclass[runningheads]{llncs}
\usepackage[T1]{fontenc}
\usepackage{graphicx,verbatim}
\usepackage{hyperref}
\usepackage{amsfonts, amsmath}
\usepackage{lipsum}
\usepackage{algorithm}
\usepackage{algpseudocode}
\usepackage{amsmath}
\usepackage{color}
\usepackage{booktabs}
\usepackage{makecell}

\usepackage{cite}

\usepackage{tikz}
\usetikzlibrary{shapes,positioning}

\begin{document}
\title{Generative Priors for Total-Body PET Reconstruction}

\author{Movindu Dassanayake\inst{1} \and 
        George Webber\inst{1} \and
        Denis Prokopenko\inst{1} \and
        Julia A. Schnabel\inst{1,2} \and
        Andrew J. Reader\inst{1}}
\authorrunning{Movindu Dassanayake et al.}
\titlerunning{Generative Priors for Total-Body PET Reconstruction}
\institute{King's College London, UK \and
           Technical University Munich, DE\\
    \email{movindu.dassanayake@kcl.ac.uk}}
  
\maketitle              

\begin{abstract}

Total-body positron emission tomography (TB-PET) enables fast, low-dose, and dynamic imaging over a long axial field of view, but reconstruction remains challenging due to reduced counts, large image volumes, and increased data dimensionality. Generative models offer powerful learned priors for low-count PET, yet explicit PET measurement conditioning at TB scale has not been demonstrated. Inspired by PET decomposed diffusion sampling (PET-DDS), we propose \emph{TB-PET-DDS}, a reconstruction framework that combines a CT-conditioned generative prior with patient-specific list-mode PET data. During reverse diffusion, PET consistency is enforced using a closed-form 3D TB-PET maximum-a-posteriori expectation maximisation fusion update, rather than relying solely on image-domain conditioning. The synthetic measurements are generated from the low-dose vendor reconstruction, providing a controlled inverse problem and retaining a full-dose reference. Proof-of-concept evaluation was performed on a single clinical TB-PET acquisition, with reconstruction hyperparameters selected against the full-dose reconstruction from the same case for all models. TB-PET-DDS improved NRMSE and SSIM relative to the DDIM and Image-DDS baselines, improving through-plane consistency in coronal views while enhancing agreement with the measured PET data. Although smoothed low-dose vendor reconstructions achieved the best global voxel-wise metrics, TB-PET-DDS produced the lowest liver coefficient of variation. These results demonstrate the feasibility of measurement conditioned diffusion reconstruction at TB scale and provide an initial route towards learned, projector-consistent generative reconstruction for low-count TB-PET imaging.


\end{abstract}

\section{Introduction}

Positron emission tomography (PET) image reconstruction is a statistically ill-posed inverse problem, particularly in the low-count regime where the measured coincidence events provide weak constraints on the unknown activity distribution \cite{Ferrara2024}. Deep learning methods have therefore been increasingly explored as learned priors for PET reconstruction \cite{reader_ai_2023,mehranian_model-based_2021,hashimoto_deep_2024}. Total-body (TB)-PET alters the reconstruction setting by providing substantially increased sensitivity over conventional scanners, enabling fast whole-body imaging, low-dose protocols, and dynamic imaging over a long axial field of view (LAFOV) \cite{MINGELS202521,Sun2024,DaubeWitherspoon2022}. However, these applications often exchange sensitivity gains for shorter acquisitions, lower injected activity, or finer temporal framing, meaning that low-count reconstruction remains a central challenge.

At the same time, TB-PET substantially increases reconstruction scale. LA-FOV systems produce larger image volumes and higher-dimensional PET data, making repeated forward and backprojection computationally demanding \cite{Prenosil2021,Dassanayake2025}. This motivates reconstruction methods that can combine strong image priors with tractable and explicit PET measurement models.

Generative models provide flexible image-domain generative priors and have recently been applied to PET reconstruction \cite{singh2024scorebased,Webber2025}. Since these priors can be trained from high-quality PET images and combined with data-consistency updates at inference time, they are attractive for low-count TB-PET. However, explicit use of PET projections and patient data with generative models have not yet been demonstrated with TB-PET.

In this work, we introduce \emph{TB-PET-DDS}, a measurement-conditioned generative reconstruction framework for TB-PET. Our contribution is threefold: first, we formulate a computationally efficient TB-scale synthetic list-mode system matrix using acquired line-of-response (LOR) and time-of-flight (TOF) information; second, we embed this system matrix inside reverse diffusion using the closed-form quadratic MAPEM update as a projector-based data-consistency step; and third, we combine this explicit PET measurement consistency with a fixed CT-conditioned pre-trained generative prior. Importantly, this PET-consistency step requires only limited additional computational headroom beyond CT-conditioned generation alone, supporting the practicality of measurement conditioned generative reconstruction at TB scale. To our knowledge, this is the first demonstration of measurement-conditioned generative reconstruction for TB-PET.

\section{Background}

\newcommand{\imx}{\mathbf{x}}
\newcommand{\imy}{\mathbf{y}}

\subsection{List-Mode PET}

Let $\imx \in \mathbb{R}_{+}^{N}$ denote the unknown activity image, and let $\mathbf{A}\in\mathbb{R}^{L \times N}$ denote a scanner-wide system matrix. Each row $\mathbf{a}_{l}^{\top}$ of $\mathbf{A}$ corresponds to one possible LOR in the scanner geometry, where $L$ is the total number of possible measurement bins. 
The likelihood contribution from the measured events depends on the corresponding rows of the system matrix, while the sensitivity term depends on the system response over the relevant measurement space. 
Accurate raw-data reconstruction therefore requires access to the scanner geometry, sensitivity, attenuation, scatter, randoms, and associated correction factors.

\subsection{List-Mode Derived Synthesised PET Model}

A practical challenge in academic TB-PET reconstruction is that the full scanner-specific reconstruction model is not always available in a usable form. We therefore use a synthesised PET acquisition framework \cite{Vass2023}. This retains measurement-domain consistency through a known forward model, and prior work has shown that such synthetic-acquisition reconstruction can outperform methods operating only in image space \cite{Dassanayake2025B,Vass2024}.

Starting from a PET image as an approximate activity distribution reconstructed by the vendor software, we generate synthetic values using an acquisition-specific LM system matrix,

\begin{equation}
\mathbf{m}_{syn}
=
\mathbf{A}_{LM}\imx_{ref}.
\label{eqn:synth_values}
\end{equation}

Here, $\mathbf{A}_{LM}\in\mathbb{R}^{M\times N}$ is the LM-derived system matrix for $M$ events,
$\mathbf{m}_{syn}\in\mathbb{R}^{M}$ contains the generated synthetic values, and $\imx_{ref}\in\mathbb{R}^{N}$ is the reference vendor reconstruction with $N$ voxels. Conceptually, $\mathbf{A}_{LM}$ is obtained by selecting the rows of the scanner-wide matrix $\mathbf{A}$ corresponding to the measured events. 

This defines a controlled reference-informed inverse problem with realistic TB-PET images and a known forward model. In this synthetic formulation, explicit line-by-line scatter, randoms, attenuation, and normalisation terms are not separately estimated during reconstruction; their effects are inherited through the vendor-corrected reference image used to generate $\mathbf{m}_{syn}$.

\subsection{PET-DDS}

PET-DDS was introduced by Singh \textit{et al.} as a PET-specific adaptation of decomposed diffusion sampling \cite{singh2024scorebased,chung2023}. A score-based generative model was first trained on synthetic PET images derived from BrainWeb dataset~\cite{collins1998design}, giving a learned prior over plausible activity distributions. At reconstruction time, sampling from this prior is interleaved with data-consistency updates derived from the PET likelihood.

At each diffusion step, the current sample is denoised to obtain Tweedie's estimate of the underlying clean image \cite{song2021scorebased}. PET-DDS then applies PET likelihood-based updates to encourage this estimate to agree with the measured data under a Poisson observation model. In this way, the generative model regularises the reconstruction, while the likelihood term conditions the sample on the PET measurements. In this work, we build on this framework for TB-PET by including LM-derived TB-PET projections in the reconstruction pipeline. 

\section{Methods}

\subsection{Clinical Data and Reference Reconstructions}
We focus on data obtained via the Siemens Biograph Vision Quadra (Quadra), a $106\,\mathrm{cm}$ LAFOV PET/CT scanner whose increased sensitivity enables low-dose, fast, and dynamic imaging, while substantially increasing reconstruction scale. The relevant data used in this work was approved via a TOTEM data request at Guy's and St Thomas' NHS Foundation Trust.

The test dataset was acquired for $420$~s following injection of $158.9$~MBq [$^{18}$F]FDG. Full-dose and low-dose vendor reconstructions were generated using the investigational software prototype (Siemens Healthineers, e7 tools). The full-dose reconstruction used all acquired LM events, whereas the low-dose reconstruction used a random $10\%$ subset of LM events. Both reconstructions were performed with $20$ expectation-maximisation (EM) iterations using vendor-specified point-spread-function (PSF) modelling.

All images were reconstructed on a grid of $440\times440\times345$ voxels with voxel size $1.65\times1.65\times3.0$~mm$^3$. The low-dose vendor reconstructions were used as the reference images $\imx_{ref}$ for generating the synthetic LM values in Eq.~\ref{eqn:synth_values}.
The acquired Quadra LM data were converted into detector-pair coordinates and TOF-bin indices (with kernel size $32$ mm), which define the acquisition-specific LM projector $\mathbf{A}_{LM}$ in ParallelProj \cite{parallelproj}.
All resulting reconstructions were evaluated against the corresponding full-dose vendor reconstructions.

\subsection{CT-conditioned Generative Prior}
To provide a patient-specific image prior, we trained a CT-conditioned score-based generative model (SGM) on static [$^{18}$F]FDG PET studies. The model follows the variance-preserving stochastic differential equation (VP-SDE) formulation of SGMs~\cite{song2021scorebased} and is trained to generate PET images from noise conditioned on the corresponding CT anatomy. A U-Net backbone~\cite{ronneberger2015unet,ho2022classifierfree} with $\sim 40M$ trainable parameters was used. Due to computational constraints, the model was trained slice-wise, with each input consisting of a single transverse PET slice and the corresponding CT slice.
The network was trained using the denoising score matching objective
\begin{equation}\label{eq:sgm-loss}
\mathcal{L}(\theta) = \mathbb{E}_{\mathbf{x}_0,t,\epsilon}
\left\|\, \sigma(t)\,\mathbf{s}_{\theta}(\mathbf{x}_t,t,\mathbf{c})+\epsilon \, \right\|_2^2,
\end{equation}
where $\epsilon \sim \mathcal{N}(0,\mathbf{I})$, $\sigma(t)$ is the marginal standard deviation, and $\mathbf{c}$ denotes the CT conditioning. 
Classifier-free guidance~\cite{ho2022classifierfree} was enabled during training by randomly zeroing the CT input channel with probability $p_{\text{uncond}}=0.4$. The value was selected empirically to provide the best balance between training convergence and generated image quality.
The training set consisted of $48$ [$^{18}$F]FDG PET studies from $47$ subjects, reconstructed using the standard clinical TOF+PSF correction chain, giving $\sim 19 K$ transverse slices on a $440 \times 440$ matrix with $1.65 \times 1.65$ mm$^2$ in-plane resolution and $3$ mm slice thickness.
The dataset was provided as a set of reconstructed images, without access to the corresponding LM data. 

This was sufficient for training the CT-conditioned SGM but did not provide the raw measurement data required for the grid-search and evaluation of the complete reconstruction framework on these studies.

During inference, the trained SGM was used as a fixed generative prior. Samples were generated using DDIM sampling~\cite{song2022denoising} with a quadratic timestep schedule, $100$ reverse diffusion steps, and $\eta=1$. We use the standard classifier-free guidance parameterisation from prior work~\cite{ho2022classifierfree, singh2024scorebased}
\begin{align}\label{eq:cfg-score}
    \mathbf{s}_{\text{cfg}}(\mathbf{x}_t,t,\mathbf{c}) = (1+\lambda_{\text{cfg}})\mathbf{s}_{\theta}(\mathbf{x}_t,t,\mathbf{c}) - \lambda_{\text{cfg}} \mathbf{s}_{\theta}(\mathbf{x}_t,t),
\end{align}
where $\lambda_{\text{cfg}}$ controls the conditioning strength and was set to $0.25$ throughout all experiments based on performance reported in~\cite{singh2024scorebased}. For volume generation and reconstruction, each reverse diffusion timestep was applied to the entire volume before proceeding to the next timestep, rather than completing the full DDIM sampling process for each slice independently. This ensured that the full volume estimate at each diffusion timestep was used by Image-DDS and TB-PET-DDS.

\subsection{LM-MLEM and Quadratic MAPEM Update}
The baseline synthetic TB-PET reconstruction was obtained using LM-MLEM \cite{Vass2023, Vass2024}. At iteration $k$, an unpenalised EM candidate image was computed as
\begin{equation}
\hat{x}_{j,\mathrm{EM}}^{(k+1)}
=
\frac{x_j^{(k)}}{s_j}
\left[
\mathbf{A}_{LM}^{\top}
\left(
\frac{\mathbf{m}_{syn}}
{\mathbf{A}_{LM}\mathbf{x}^{(k)}+\epsilon}
\right)
\right]_j,
\qquad
s_j =
\left[\mathbf{A}_{LM}^{\top}\mathbf{1}\right]_j ,
\label{eqn:lm_mlem_candidate}
\end{equation}
where $\epsilon$ is a small numerical constant and $\mathbf{s}$ is the sensitivity image defined with respect to the synthetic LM geometry used in this work. The sensitivity is therefore acquisition-specific, with the rows of $\mathbf{A}_{LM}$ determined by the measured LOR and TOF information.

For penalised reconstruction, we incorporate prior information through a quadratic penalty within the MAPEM framework~\cite{Green1990,DePierro1995,Fessler1995}. Given a prior or anchor image $\mathbf{z}^{(k)}$ at iteration $k$, the penalised update for voxel $j$ is obtained by minimising the EM surrogate with a quadratic fusion term
$
\frac{\beta}{2}
\left(
x_j-z_j^{(k)}
\right)^2 ,
$
where $\beta$ controls the strength of fusion to $\mathbf{z}^{(k)}$. This gives the closed-form update
\begin{equation}
x_j^{(k+1)}
=
\frac{1}{2}
\left[
z_j^{(k)} - \frac{s_j}{\beta}
\right]
+
\frac{1}{2}
\sqrt{
\left(
z_j^{(k)} - \frac{s_j}{\beta}
\right)^2
+
4\hat{x}_{j,\mathrm{EM}}^{(k+1)}
\frac{s_j}{\beta}
}.
\label{eqn:quadratic_fusion_mapem}
\end{equation}
For small $\beta$, the update approaches the unpenalised LM-MLEM candidate, while larger $\beta$ increases fusion towards $\mathbf{z}^{(k)}$.

\subsection{TB-PET-DDS}

Our proposed method \emph{TB-PET-DDS}, adapted from DDIP \cite{Hashimoto2026}, combines a fixed CT-conditioned generative prior with the synthetic LM PET measurement model, and is given in Algorithm~\ref{alg:pet_dds}. Unlike DDIP-based PET reconstruction, the score model is not fine-tuned at inference time; measurement consistency is imposed only through the PET reconstruction update. Compared with PET-DDS \cite{singh2024scorebased}, we replace explicit likelihood gradient steps with the synthetic LM MAPEM update in Eq.~\ref{eqn:quadratic_fusion_mapem}. This gives a closed-form proximal update, with the generative estimate acting as the quadratic anchor.

At each reverse diffusion step, the score model freely generates PET slices from the current noisy image and CT. These slices then form an anchor image volume, updated using the LM MAPEM step, and then used as the clean-image estimate in the DDIM reverse update.

\begin{algorithm}[t]
\caption{TB-PET-DDS reconstruction with synthetic LM measurement conditioning. Adapted from \cite{Hashimoto2026}.}
\label{alg:pet_dds}
\begin{algorithmic}[1]
\Statex \textbf{Input:} synthetic LM data $\mathbf{m}_{syn}$, LM projector $\mathbf{A}_{LM}$, sensitivity image $\mathbf{s}$
\Statex \textbf{Input:} low-dose PET image $\mathbf{x}_{LD}$, CT image $\mathbf{g}$, trained score model $\epsilon_{\theta}$
\Statex \textbf{Input:} diffusion schedule $T',\ldots,1$, inner MAPEM iterations $M$, fusion weight $\lambda_{\mathrm{DDS}}$
\Statex \textbf{Output:} reconstructed PET image $\hat{\mathbf{x}}$

\State Initialise $\mathbf{x}_{T'}$ by adding diffusion noise to $\mathbf{x}_{LD}$

\For{$t = T',\ldots,1$}

    \State Predict PET slices from $\mathbf{x}_{t}$ using the CT-conditioned score model $\epsilon_{\theta}(\cdot,t,\mathbf{g})$
    \State Stack the predicted slices to obtain a 3D generative anchor image $\mathbf{z}^{(t)}$
    \State Enforce non-negativity on $\mathbf{z}^{(t)}$

    \State Set $\mathbf{x}^{(t,0)} \gets \mathbf{z}^{(t)}$

    \For{$m = 0,\ldots,M-1$}
        \State Compute the LM-MLEM candidate $\hat{\mathbf{x}}_{\mathrm{EM}}^{(t,m+1)}$ using $\mathbf{m}_{syn}$, $\mathbf{A}_{LM}$ and $\mathbf{s}$
        \State Fuse $\hat{\mathbf{x}}_{\mathrm{EM}}^{(t,m+1)}$ with $\mathbf{z}^{(t)}$ using Eq.~\ref{eqn:quadratic_fusion_mapem}, with $\beta=\lambda_{\mathrm{DDS}}$
        \State Store the fused result as $\mathbf{x}^{(t,m+1)}$
    \EndFor

    \State Set the data-consistent estimate $\hat{\mathbf{x}}_{0}^{(t)} \gets \mathbf{x}^{(t,M)}$
    \State Use $\hat{\mathbf{x}}_{0}^{(t)}$ in the DDIM reverse update to obtain $\mathbf{x}_{t-1}$

\EndFor

\State \Return $\hat{\mathbf{x}} \gets \hat{\mathbf{x}}_{0}^{(1)}$
\end{algorithmic}
\end{algorithm}

\subsection{Baselines and Metrics}

We compare TB-PET-DDS against conventional, post-processing, and generative-based baselines. The \emph{Low-dose vendor} baseline is the vendor reconstruction obtained from the same $10\%$ LM event subset used for the low-count setting. The \emph{Synthetic LM-MLEM} baseline reconstructs the synthetic LM measurements using Eq.~\ref{eqn:lm_mlem_candidate}. The \emph{Low-dose + PS} baseline applies isotropic Gaussian post-smoothing to the low-dose vendor reconstruction. The smoothing width is selected by grid search to minimise NRMSE against the full-dose reference, providing an upper-bound post-smoothing baseline.

We also compare against two generative baselines. \emph{DDIM} uses the CT-conditioned score model without PET measurement consistency, generating PET images from the learned prior and CT conditioning alone. \emph{Image-DDS} applies DDS conditioning directly on the low-dose PET image using an image-domain squared-error update, rather than using the PET projector:

\begin{equation}
\imx^{t,m+1}
=
\imx^{t,m}
-
2\alpha_t
[
\lambda_{DDS}(\imx^{t,m}-\imx_{LD})
+
(1-\lambda_{DDS})(\imx^{t,m}-\imx_{anchor}^{t}).
\label{eqn:Image-DDS}
\end{equation}

For this exploratory single-case study, we performed a grid search over the number of DDS steps, the number of inner MAPEM iterations, and the fusion weight $\lambda_{\mathrm{DDS}}$ for both Image-DDS and TB-PET-DDS. Hyperparameters were selected using normalised root-mean-square error (NRMSE) against the full-dose vendor reconstruction and then fixed for the comparisons. 
Consequently, the reported results represent same-case hyperparameter optimisation rather than performance following independent validation, and should therefore be interpreted as proof-of-concept results rather than estimates of generalisable reconstruction performance.

Reconstructions were evaluated against the full-dose vendor reconstruction $\mathbf{x}_{\mathrm{FD}}$ after mean normalisation. We report NRMSE, structural similarity index measure (SSIM), liver coefficient of variation (CoV), reconstruction time, and peak memory usage. Liver CoV provides an exploratory measure of relative variation within an approximately uniform region. However, CoV may also be reduced by smoothing and therefore cannot, in isolation, establish improved spatial resolution or overall image quality.

\section{Results}

\subsection{Quantitative Comparison}

\begin{table*}[t!]
\centering
\caption{Quantitative comparison of low-dose and generative-based PET reconstruction methods against the full-dose vendor reconstruction. All reported values are using the full mean-normalised volumes. }
\label{tab:quantitative_comparison}
\scriptsize
\setlength{\tabcolsep}{4pt}
\begin{tabular}{lccccccc}
\toprule
\textbf{Metric} &
\makecell{\textbf{Low-dose}\\\textbf{vendor}} &
\makecell{\textbf{Synthetic}\\\textbf{MLEM}} &
\makecell{\textbf{Low-dose}\\\textbf{+ PS}} &
\makecell{\textbf{DDIM}\\\textbf{CT-cond.}} &
\makecell{\textbf{Image}\\\textbf{DDS}} &
\makecell{\textbf{TB-PET}\\\textbf{DDS}} \\
\midrule
NRMSE (\%)        $\downarrow$   & 16.32 & 15.08 & 12.08 & 56.92 & 56.44 & 27.93 \\
SSIM              $\uparrow$     & 0.991 & 0.992 & 0.995 & 0.875 & 0.913 & 0.971 \\
Liver CoV (\%)                   & 21.24 & 18.93 & 13.50 & 15.59 & 9.67 & 8.98 \\
Recon. time (min) $\downarrow$   & --    & 9.71  & --    & 80.60 & 86.20 & 101.17 \\
Peak Memory (GB)  $\downarrow$   & --    & 12    & --    & 16    & 16    & 16    \\
\bottomrule
\end{tabular}
\end{table*}

Quantitative results are reported in Table~\ref{tab:quantitative_comparison}. Low-dose + PS perform the best throughout for NRMSE and SSIM, followed by Low-dose vendor and then the proposed method. The proposed method outperforms the other generative methods. We also measured CoV to quantify liver uniformity, with the selected regions shown in the green box in Fig.~\ref{fig:TB-PET recons} (c). TB-PET-DDS also produced the lowest liver CoV in the selected ROI at 8.98\%. Lastly, when comparing the slice-wise NRMSE, the proposed method contains some slices that have a lower NRMSE than the baselines, as indicated within Fig~\ref{fig:TB-PET recons}. 

Runtime and memory usage are also reported in Table~\ref{tab:quantitative_comparison}. TB-PET-DDS adds a 3D LM MAPEM update at each diffusion step, increasing runtime relative to image-only DDS. However, the update is closed-form and does not require adapting the score model at inference time. This enables practical TB-PET reconstruction with explicit PET conditioning.

\subsection{Qualitative Comparison}

\begin{figure}[t!]
\centering
\includegraphics[width=1\linewidth]{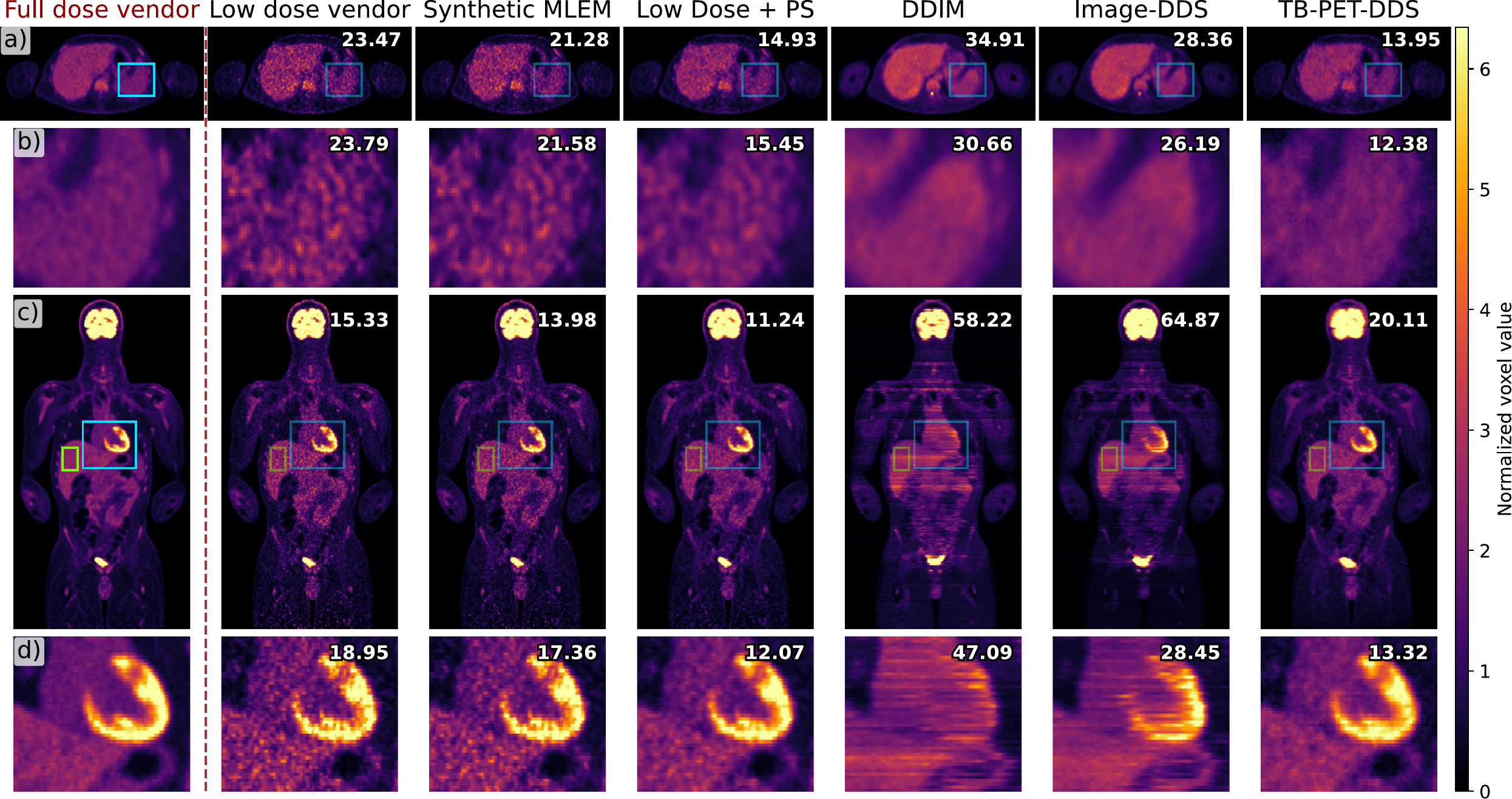}
\caption{TB-PET reconstructions with transaxial views (a) with a zoom insert (b), and coronal views (c) with a zoom insert (d). Zoom insert ROI are indicated in blue and the liver CoV ROI is indicated in green. Slice-wise NRMSE against the full-dose reconstruction (the reference, shown in red) is embedded within the image.}
\label{fig:TB-PET recons}
\end{figure}

Representative reconstructions are shown in Fig.~\ref{fig:TB-PET recons}. DDIM provides a baseline of the learned anatomical prior without PET data, while Image-DDS tests whether image-domain conditioning on the low-dose vendor reconstruction is sufficient. TB-PET-DDS instead uses the synthetic LM projector during sampling, allowing PET measurement consistency to be imposed in 3D.

With the introduction of the SGM, we can see that the images become smoother but lose detail in critical regions, which can be seen in the axial views in Fig~\ref{fig:TB-PET recons} (a), (b). While Image-DDS recovers some information from the low-dose image, as seen on the coronal zoom insert in Fig~\ref{fig:TB-PET recons} (d), there are major artifacts in this view. This is due to the slice-wise image generation from the SGM which has little slice-consistency. TB-PET-DDS shows reduced through-plane artefacts relative to the image-domain generative baselines; there is stronger slice consistency (due to the 3D projector) and heavy noise reduction due to the SGM prior. However, there remain some regions throughout the TB-PET image where there is still some slice inconsistency ,and since no dedicated through-plane consistency metric or 2D/3D ablation is included, these observations remain qualitative.

\section{Discussion and Conclusions}
This work presents a proof-of-concept application of projector-conditioned generative reconstruction to TB-PET, combining a CT-conditioned generative prior with a synthetic LM PET model and closed-form MAPEM conditioning. On the single evaluation case, TB-PET-DDS improved NRMSE and SSIM relative to DDIM and Image-DDS, but remained worse than post-smoothed low-dose reconstruction on global metrics. TB-PET-DDS also produced the lowest liver CoV, although this may reflect increased smoothing and cannot alone demonstrate improved regional image quality. Qualitatively, the 3D PET data term reduced through-plane inconsistencies introduced by the 2D slice-wise generative prior.

The principal limitations are the single-case evaluation and hyperparameter selection using the same case as the final comparison, such that the reported results do not represent true held-out validation. PET-DDS was also not directly evaluated, preventing conclusions regarding the relative performance of MAPEM and likelihood-gradient conditioning. The current 2D prior additionally motivates future work using slice-aware or volumetric generative models and evaluation across multiple subjects with resolution- or task-sensitive metrics.

Overall, these results demonstrate the feasibility of integrating a generative prior with 3D projector-based PET consistency at TB scale, representing an initial application of this approach rather than evidence of generalisable superiority over existing reconstruction methods.

\bibliography{art.bib}

@article{Hashimoto2026,
  author   = {Hashimoto, Fumio and Gong, Kuang},
  title    = {{PET} Image Reconstruction Using {Deep Diffusion Image Prior}},
  journal  = {IEEE Transactions on Medical Imaging},
  year     = {2026},
  volume   = {45},
  number   = {6},
  pages    = {2628--2638},
  doi      = {10.1109/TMI.2026.3659792}
}

@article{Webber2025,
  author    = {Webber, George and Mizuno, Yuya and Howes, Oliver D. and Hammers, Alexander and King, Andrew P. and Reader, Andrew J.},
  title     = {Likelihood-Scheduled {Score-Based Generative Modeling} for Fully {3D} {PET} Image Reconstruction},
  journal   = {IEEE Transactions on Medical Imaging},
  publisher = {Institute of Electrical and Electronics Engineers (IEEE)},
  year      = {2025},
  volume    = {44},
  number    = {11},
  pages     = {4445--4456},
  issn      = {1558-254X},
  doi       = {10.1109/tmi.2025.3576483},
}

@article{MINGELS202521,
  author   = {Clemens Mingels and Kevin J. Chung and Austin R. Pantel and Axel Rominger and Ian Alberts and Benjamin A. Spencer and Lorenzo Nardo and Thomas Pyka},
  title    = {{Total-Body} {PET/CT}: Challenges and Opportunities},
  journal  = {Seminars in Nuclear Medicine},
  year     = {2025},
  volume   = {55},
  number   = {1},
  pages    = {21--30},
  issn     = {0001-2998},
  doi      = {10.1053/j.semnuclmed.2024.08.003}
}

@article{Ferrara2024,
  author    = {Ferrara, Daria and Shiyam Sundar, Lalith Kumar and Chalampalakis, Zacharias and Geist, Barbara Katharina and Gompelmann, Daniela and Gutschmayer, Sebastian and Hacker, Marcus and Kertész, Hunor and Kluge, Kilian and Idzko, Marco and Langsteger, Werner and Yu, Josef and Rausch, Ivo and Beyer, Thomas},
  title     = {Low-dose and standard-dose {Whole-Body} {[18F]FDG-PET/CT} imaging: implications for healthy controls and lung cancer patients},
  journal   = {Frontiers in Physics},
  publisher = {Frontiers Media SA},
  year      = {2024},
  month     = {Apr},
  volume    = {12},
  issn      = {2296-424X},
  doi       = {10.3389/fphy.2024.1378521},
}

@article{Sun2024,
  author    = {Sun, Yuanyuan and Cheng, Zhaoping and Qiu, Jianfeng and Lu, Weizhao},
  title     = {Performance and application of the {Total-Body} {PET/CT} scanner: a literature review},
  journal   = {EJNMMI Research},
  publisher = {Springer Science and Business Media LLC},
  year      = {2024},
  volume    = {14},
  number    = {1},
  issn      = {2191-219X},
  doi       = {10.1186/s13550-023-01059-1},
}

@article{DaubeWitherspoon2022,
  author    = {Daube-Witherspoon, Margaret E and Pantel, Austin R and Pryma, Daniel A and Karp, Joel S},
  title     = {{Total-Body} {PET}: a new paradigm for molecular imaging},
  journal   = {The British Journal of Radiology},
  publisher = {Oxford University Press (OUP)},
  year      = {2022},
  month     = {Dec},
  volume    = {95},
  number    = {1140},
  issn      = {1748-880X},
  doi       = {10.1259/bjr.20220357},
}

@article{Green1990,
  author  = {Green, Peter J.},
  title   = {Bayesian Reconstructions from Emission Tomography Data Using a Modified {EM} Algorithm},
  journal = {IEEE Transactions on Medical Imaging},
  year    = {1990},
  volume  = {9},
  number  = {1},
  pages   = {84--93},
  doi     = {10.1109/42.52985}
}

@article{DePierro1995,
  author  = {De Pierro, Alvaro R.},
  title   = {A Modified {Expectation Maximization} Algorithm for Penalized Likelihood Estimation in Emission Tomography},
  journal = {IEEE Transactions on Medical Imaging},
  year    = {1995},
  volume  = {14},
  number  = {1},
  pages   = {132--137},
  doi     = {10.1109/42.370409}
}

@article{Fessler1995,
  author  = {Fessler, Jeffrey A. and Hero, Alfred O.},
  title   = {Penalized {Maximum-Likelihood} Image Reconstruction Using Space-Alternating Generalized {EM} Algorithms},
  journal = {IEEE Transactions on Image Processing},
  year    = {1995},
  volume  = {4},
  number  = {10},
  pages   = {1417--1429},
  doi     = {10.1109/83.465106}
}

@inproceedings{
chung2023,
title={Decomposed Diffusion Sampler for Accelerating Large-Scale Inverse Problems},
author={Hyungjin Chung and Suhyeon Lee and Jong Chul Ye},
booktitle={The Twelfth International Conference on Learning Representations},
year={2024},
doi = {10.48550/arXiv.2303.05754}
}

@article{Prenosil2021,
  author    = {Prenosil, George A. and Sari, Hasan and F\"{u}rstner, Markus and Afshar-Oromieh, Ali and Shi, Kuangyu and Rominger, Axel and Hentschel, Michael},
  title     = {Performance Characteristics of the {Biograph Vision Quadra} {PET/CT} System with a Long Axial Field of View Using the {NEMA NU 2-2018} Standard},
  journal   = {Journal of Nuclear Medicine},
  publisher = {Society of Nuclear Medicine},
  year      = {2021},
  volume    = {63},
  number    = {3},
  pages     = {476--484},
  issn      = {2159-662X},
  doi       = {10.2967/jnumed.121.261972},

}

@article{Vass2023,
  author={Vass, Laurence and Reader, Andrew J.},
  journal={IEEE Transactions on Radiation and Plasma Medical Sciences}, 
  title={Synthesized Image Reconstruction for Post-Reconstruction Resolution Recovery}, 
  year={2023},
  volume={7},
  number={5},
  pages={473-482},
  doi={10.1109/TRPMS.2023.3247489}}

@article{Dassanayake2025,
  author    = {Dassanayake, Movindu and Lopez, Alejandro and Reader, A. J. and Cook, Gary J.R. and Mingels, Clemens and Rahmim, Arman and Seifert, Robert and Alberts, Ian and Yousefirizi, Fereshteh},
  title     = {Artificial Intelligence-Guided {PET} Image Reconstruction and Multi-Tracer Imaging},
  journal   = {PET Clinics},
  publisher = {Elsevier BV},
  year      = {2025},
  month     = {oct},
  volume    = {20},
  number    = {4},
  pages     = {453--461},
  issn      = {1556-8598},
  doi       = {10.1016/j.cpet.2025.07.005},
}

@inproceedings{Vass2024,
  author    = {Vass, Laszlo and Reader, Andrew J.},
  title     = {Synthesised {PET} Image Reconstruction with Kernel Methods and the {Deep Image Prior} for Post-Reconstruction Image Enhancement},
  booktitle = {2024 IEEE Nuclear Science Symposium (NSS), Medical Imaging Conference (MIC) and Room Temperature Semiconductor Detector Conference (RTSD)},
  publisher = {IEEE},
  year      = {2024},
  pages     = {1--1},
  doi       = {10.1109/NSS/MIC/RTSD57108.2024.10658573},
}

@article{parallelproj,
  author    = {Schramm, Georg and Thielemans, Kris},
  title     = {{PARALLELPROJ}---An {Open-Source} Framework for Fast Calculation of Projections in Tomography},
  journal   = {Frontiers in Nuclear Medicine},
  publisher = {Frontiers Media SA},
  year      = {2024},
  month     = {jan},
  issn      = {2673-8880},
  doi       = {10.3389/fnume.2023.1324562},
}

@article{reader_ai_2023,
  author = {Reader, Andrew J and Pan, Bolin},
  title    = {{AI} for {PET} image reconstruction},
  journal  = {British Journal of Radiology},
  year     = {2023},
  volume   = {96},
  number   = {1150},
  pages    = {20230292},
  issn     = {0007-1285},
  doi      = {10.1259/bjr.20230292},
  urldate  = {2025-02-17},
  pmid     = {37486607},
  pmcid    = {PMC10546435}
}

@article{mehranian_model-based_2021,
  author   = {Mehranian, Abolfazl and Reader, Andrew J.},
  title    = {Model-Based {Deep Learning} {PET} Image Reconstruction Using {Forward--Backward Splitting} {Expectation--Maximization}},
  journal  = {IEEE Transactions on Radiation and Plasma Medical Sciences},
  year     = {2021},
  volume   = {5},
  number   = {1},
  pages    = {54--64},
  issn     = {2469-7303},
  doi      = {10.1109/TRPMS.2020.3004408},
  urldate  = {2025-03-10},
  pmid     = {34056150},
  pmcid    = {PMC7610859}
}

@article{hashimoto_deep_2024,
  author     = {Hashimoto, Fumio and Onishi, Yuya and Ote, Kibo and Tashima, Hideaki and Reader, Andrew J. and Yamaya, Taiga},
  title      = {{Deep Learning}-based {PET} image denoising and reconstruction: a review},
  journal    = {Radiological Physics and Technology},
  year       = {2024},
  volume     = {17},
  number     = {1},
  pages      = {24--46},
  shorttitle = {Deep learning-based {PET} image denoising and reconstruction},
  issn       = {1865-0341},
  doi        = {10.1007/s12194-024-00780-3},
  urldate    = {2025-02-18},
  pmid       = {38319563},
  pmcid      = {PMC10902118},
  language   = {en}
}

@misc{song2021scorebased,
  author        = {Song, Yang and {Sohl-Dickstein}, Jascha and Kingma, Diederik P. and Kumar, Abhishek and Ermon, Stefano and Poole, Ben},
  title         = {{Score-Based Generative Modeling} through {Stochastic Differential Equations}},
  publisher     = {arXiv},
  year          = {2021},
  month         = {feb},
  number        = {arXiv:2011.13456},
  doi           = {10.48550/arXiv.2011.13456},
  urldate       = {2025-07-31},
  eprint        = {2011.13456},
  primaryclass  = {cs},
  archiveprefix = {arXiv}
}

@article{singh2024scorebased,
  author        = {Singh, Imraj RD and Denker, Alexander and Barbano, Riccardo and Kereta, {\v Z}eljko and Jin, Bangti and Thielemans, Kris and Maass, Peter and Arridge, Simon},
  title         = {{Score-Based Generative Models} for {PET} Image Reconstruction},
  journal       = {Machine Learning for Biomedical Imaging},
  year          = {2024},
  volume        = {2},
  number        = {Generative Models},
  pages         = {547--585},
  issn          = {2766-905X},
  doi           = {10.59275/j.melba.2024-5d51},
  urldate       = {2025-07-24},
  eprint        = {2308.14190},
  primaryclass  = {eess},
  archiveprefix = {arXiv}
}

@inproceedings{ronneberger2015unet,
  author       = {Ronneberger, Olaf and Fischer, Philipp and Brox, Thomas},
  title        = {{U-Net}: Convolutional networks for biomedical image segmentation},
  booktitle    = {International Conference on Medical image computing and computer-assisted intervention},
  year         = {2015},
  pages        = {234--241},
  organization = {Springer},
  doi          = {10.1007/978-3-319-24574-4_28}
}

@inproceedings{song2022denoising,
 author = {Ho, Jonathan and Jain, Ajay and Abbeel, Pieter},
 booktitle = {Advances in Neural Information Processing Systems},
 editor = {H. Larochelle and M. Ranzato and R. Hadsell and M.F. Balcan and H. Lin},
 pages = {6840--6851},
 publisher = {Curran Associates, Inc.},
 title = {{Denoising Diffusion Probabilistic Models}},
 doi = {10.48550/arXiv.2006.11239},
 volume = {33},
 year = {2020}
}

@inproceedings{
ho2022classifierfree,
title={{Classifier-Free Diffusion Guidance}},
author={Jonathan Ho and Tim Salimans},
booktitle={NeurIPS 2021 Workshop on Deep Generative Models and Downstream Applications},
year={2021},
doi={10.48550/arXiv.2207.12598}
}

@inproceedings{Dassanayake2025b,
  title = {{Region of Interest Based Fast High Quality Time-Of-Flight List-Mode Pet Reconstruction}},
  DOI = {10.1109/nss/mic/rtsd57106.2025.11287285},
  booktitle = {2025 IEEE Nuclear Science Symposium (NSS),  Medical Imaging Conference (MIC) and Room Temperature Semiconductor Detector Conference (RTSD)},
  publisher = {IEEE},
  author = {Dassanayake,  M. and Schnabel,  J. A. and Reader,  A. J.},
  year = {2025},
  pages = {1–1}
}

@article{collins1998design,
  title={{Design and construction of a realistic digital brain phantom}},
  author={Collins, D Louis and Zijdenbos, Alex P and Kollokian, Vasken and Sled, John G and Kabani, Noor Jehan and Holmes, Colin J and Evans, Alan C},
  journal={{IEEE Transactions on Medical Imaging}},
  volume={17},
  number={3},
  pages={463--468},
  year={1998},
  publisher={IEEE},
  doi={10.1109/42.712135}
}

\bibliographystyle{splncs04}

\end{document}